\documentclass[a4paper,11pt,twocolumn]{article}

\usepackage[dvips]{graphicx}
\usepackage{amsmath}
\usepackage{mathtext}
\usepackage{cite}

\begin{document}

\title{Local Diamagnetic Susceptibility of Quasi-Two-Dimensional Graphite}

\author{\bf \em E. G. Nikolaev $^a$\thanks{e-mail: nikolaev@kapitza.ras.ru},
A. S. Kotosonov $^b$,\\
\bf \em E. A. Shalashugina $^a$, A. M. Troyanovskii $^a$,\\
\bf \em and V. I. Tsebro  $^c$,\\
{\small \em $^a$Kapitsa Institute for Physical Problems, Russian Academy of Sciences,}\\
{\small \em ul. Kosygina 2, Moscow, 119334 Russia}\\
{\small \em $^b$OAO NIIgrafit, ul. Electrodnaya 2, Moscow, 111524, Russia}\\
{\small \em $^c$Lebedev Physical Institute, Russian Academy of Sciences,}\\
{\small \em Leninskii pr. 53, Moscow, 119991 Russia}\\
}

\date{February, 20, 2013}

\maketitle

\begin{abstract}
A sample of quasi-two-dimensional graphite (QTDG) whose magnetic properties are described within the Dirac fermion model is investigated by the nuclear magnetic resonance (NMR) and scanning tunneling microscopy (STM) techniques. The broad spectrum of the sample points to a large dispersion of crystallite sizes in this system, which is also confirmed by STM data. It is established that the local diamagnetic susceptibility may substantially exceed the average value over the sample and reaches an abnormally high value of $-1.3\times 10^{-4}$ emu/g at $T = 4.2$~K, which is greater than the corresponding value of highly oriented graphite by a factor of four.

\vspace{36pt}
%PACS 73.20.Dx, 73.50.Jt

\end{abstract}

Today there is a large number of publications on the study of transport properties of graphene (see, for example, the review \cite{sarma11}); however, there is quite limited experimental information on the magnetic properties of graphene. This is associated, first of all, with the extremely small mass of a graphene sheet, which complicates the direct measurement of its magnetization, as well as with other factors. For example, in \cite{sepioni10}, the authors measured the magnetization of a sample consisting of a large number of isolated graphene sheets by a SQUID magnetometer; however, they failed to correctly determine the proper diamagnetic susceptibility of graphene at low temperatures because of a large paramagnetic contribution of localized magnetic moments. The application of local techniques, such as NMR, is hampered, in addition to the sensitivity problem, by the fact that the demagnetization factor for a magnetic field directed perpendicular to the plane of a sheet for such a plane object as the graphene sheet is equal to unity, which leads to the full compensation of the sample magnetization contribution to the local field. At the same time, for a rather long time, there has been known a carbon material such as quasi-two-dimensional graphite (QTDG) \cite{kotosonov84,kotosonov86}, which exists in the form of large bulk samples and provides insight into the magnetic properties of graphene. In \cite{kotosonov86}, the author established that the susceptibility of the best samples of QTDG along the $c$ axis amounts to $-7.5\times 10^{-5}$~emu/g at $T = 4.2$~K, which is a record value after superconductors. In this paper, the author also showed that the temperature dependence of $\chi_c$ in a wide range of temperatures can be rather well described by the formula for an isolated graphene plane, which was first proposed by McClure \cite{mcclure56} and was modified with regard to the broadening of the linear energy spectrum near the Dirac point due to the electron scattering by structural defects.

In contrast to single-crystal graphite and highly oriented graphite, QTDG shows no azimuthal ordering of layers along the $c$ axis \cite{kotosonov84,kotosonov86}. The important role of the mutual orientation of carbon layers was confirmed by the calculations of the energy-band structure of multilayer graphene \cite{santos56,latil07}. In these works, the authors showed that, in the absence of azimuthal ordering of layers, the samples also exhibit a linear dispersion law, which is characteristic of Dirac massless fermions and is responsible for a whole series of anomalous properties of graphene, including the behavior of susceptibility considered in \cite{mcclure56}. In \cite{ominato12}, it was pointed out that a system consisting of defect-free misaligned graphene layers should exhibit nearly perfect diamagnetism at low temperatures. However, one should note that, in the case of QTDG, the presence of structural defects is essential, because it is these defects that allow one to realize a structure in which there is no correlation between adjacent carbon planes. However, the presence of a large number of defects restricts the size of crystallites in QTDG, which, in turn, limits the value of magnetic susceptibility.

For an inhomogeneous system such as QTDG, it is of interest to study, along with the mean values, the variations of such parameters as the crystallite size and the local magnetic susceptibility and, on the basis of these data, to evaluate how far the local susceptibility can exceed its mean value over a sample. To this end, we applied two methods in the present study: NMR, which enables one to determine the dispersion of local magnetization in the system, and STM, which provides direct information on the size of crystallites in QTDG.

The QTDG was obtained by depositing the products of pyrolysis of hydrocarbons at $T = 2100^\circ $C onto a plane substrate and represented a polycrystalline plate on which crystallites are arranged so that the carbon layers are predominantly parallel to the plane of the substrate, and, hence, the axis of the texture coincides with the normal to the plane of the substrate. The distance between carbon planes in the sample, determined by the position of the (002) X-ray reflex, is 3.422(1)~{\AA}, which is much greater than the same distance in a highly oriented graphite (3.3601(1)~{\AA}) and points to the absence of correlation between carbon layers \cite{kotosonov84}. The X-ray diffraction analysis has also shown that the texture parameter $\langle \sin^2\Theta\rangle$, which characterizes the misalignment of crystallites, amounts to 0.097 ($\Theta$ is the angle between the $c$ axis of a crystallite and the axis of the texture). The measurement of the magnetization of the sample by a ballistic magnetometer with regard to the textural correction $1 - \langle \sin^2\Theta\rangle$ yielded a value of $\chi_c  = -(5.63 \pm 0.05) \times 10^5$~emu/g for susceptibility at $T = 4.2$~K. For comparison, the measurement of the susceptibility of a highly oriented graphite for $H \|\,c$ yielded a value of $ -(3.29 \pm 0.03) \times 10^5$~emu/g at $T = 4.2$~K.

\begin{figure}[h]
\begin{center}
\includegraphics[width=14cm]{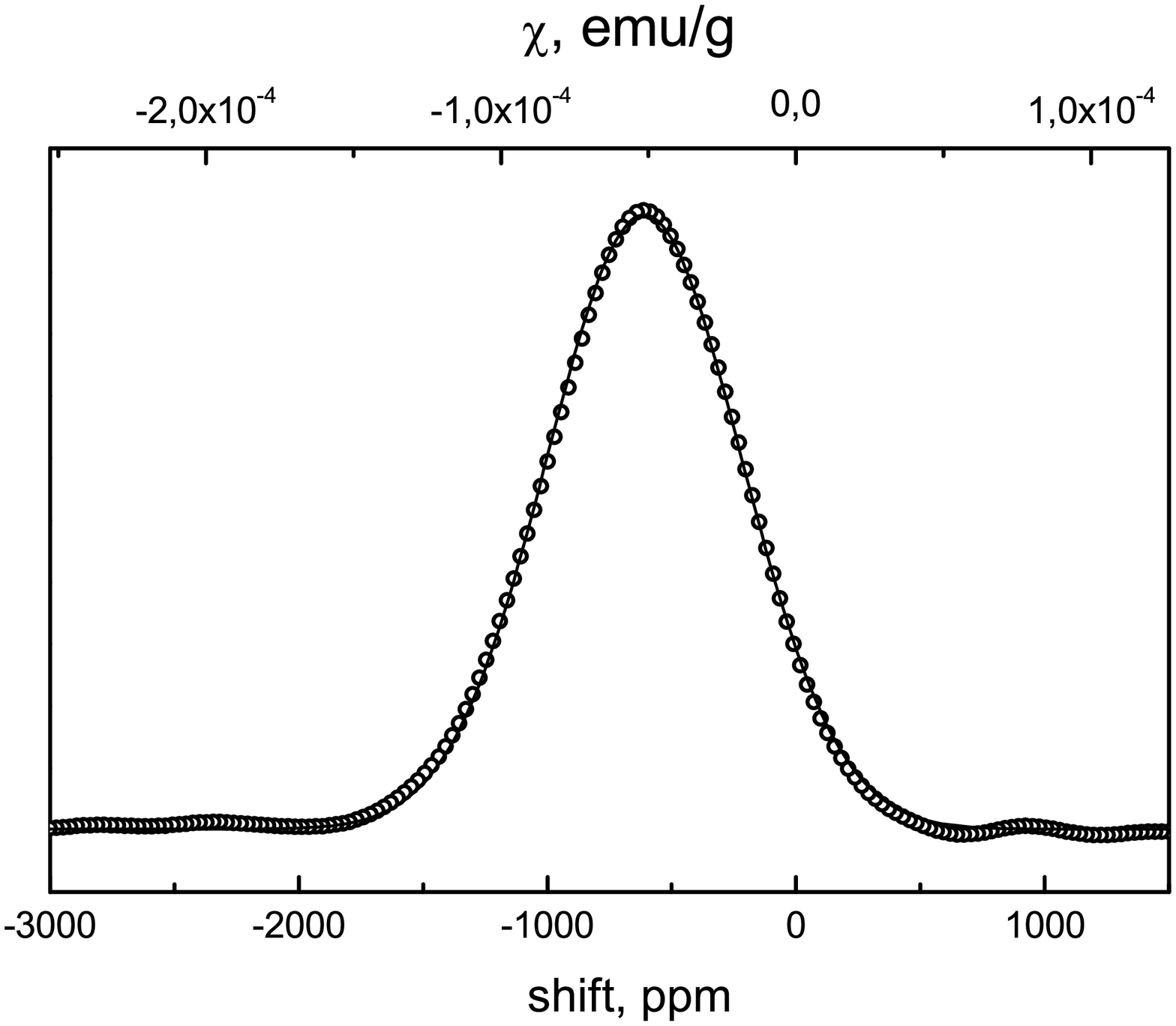}
  \caption{NMR spectrum of quasi-two-dimensional graphite at $T$ = 4.2 K and $H$ = 21 kOe; the dots correspond to experiment, and the solid line represents an approximation by a Gaussian.}
  \label{fig:MNR_spectra}
\end{center}
\end{figure}

A sample of QTDG for the NMR analysis was cut out in the form of a long cylinder (the length 13~mm and the diameter 4~mm) with the axis perpendicular to the texture axis. Figure \ref{fig:MNR_spectra} shows the $^{13}$C NMR spectrum (with natural concentration of 1.1\%) of the QTDG sample for the magnetic field parallel to the texture axis ($T = 4.2$~K and $H = 21$~kOe). In graphite for the magnetic field perpendicular to carbon layers ($H \|\,c$) and in the absence of demagnetization effects the only significant contribution to the shift is associated with the Lorentz field due to the macroscopic magnetization of the medium, which is assumed to coincide with magnetic induction in the graphite structure \cite{goze02}. In this case, with regard to the demagnetization factor, the shift is expressed as $\delta_c = 4\pi(1-D)\chi_c\rho$ ($\rho = 2.2$~g/cm$^3$ is the density of the sample). For a long cylinder in a field perpendicular to its axis, $D = 1/2$; therefore, the formula for calculating susceptibility from the shift reduces to $\chi_c = \delta_c/2\pi \rho$. The value of $\chi_c$ obtained from the position of the maximum of the spectrum with the use of this formula and with regard to the textural correction amounts to $-(5.0 \pm 0.2)\times 10^5$~emu/g. The small difference from the result of direct measurement of susceptibility given above is likely to be associated with the fact that the Lorentz field for the graphite structure is nevertheless slightly less than magnetic induction. The upper scale in Fig.~\ref{fig:MNR_spectra} is graduated in values of $\chi$ calculated by the above formula with regard to a correction factor equal to the ratio of the values of susceptibility obtained by the ballistic method to those obtained from the NMR spectrum.

The misalignment of crystallites with respect to the texture axis should make contribution to the broadening of the NMR spectrum in QTDG, because the susceptibility of graphite strongly depends on the orientation of carbon layers with respect to the magnetic field. However, the evaluation of this contribution with the use of the texture parameter yields a value of 39~ppm, which is much less than the observed width, amounting to 920~ppm at the half-maximum. Note that the spectrum of the QTDG sample has a Gaussian shape to a good degree of accuracy (Fig.~\ref{fig:MNR_spectra}), which can be considered as an indicative of the statistical nature of the broadening. It is natural to associate this broadening with the dispersion of crystallite sizes in the sample. According to \cite{kotosonov84}, experimental results on the measurement of transport properties in QTDG can be reasonably interpreted under the assumption that the main source of carriers are the dislocation boundaries of two-dimensional crystallites, with one hole carrier per crystallite. Thus, there is a single valued dependence between the mean size of crystallites $L$ in the plane and the Fermi temperature $T_{F0}$, determined by the number of carriers in the system; this dependence is expressed by the formula $L = 1.48 \times 10^{-4}/T_{F0}$~nm \cite{kotosonov91}. The relationship between $T_{F0}$ and the diamagnetic susceptibility of QTDG can be described by the formulas that were used in \cite{kotosonov86} for approximating the temperature dependence of $\chi_c$:
\begin{equation}\label{eq:chi}
    \begin{split}
    & \qquad \chi = -1.46 \times 10^{-3}\gamma^2_0 \ \times \\
    & \times {\text{sech}}^2(T_F/2(T+\Delta))/(T+\Delta) \ \text{emu/g}
    \end{split}
\end{equation}
where $\gamma_0 = 3$~eV is the energy band parameter, $T_F$ is the chemical potential in units of temperature, and $\Delta \approx T_{F0}/2$ is a parameter that takes into account the broadening of the spectrum due to the scattering by impurities. For every temperature, $T_F$ is determined from the electric neutrality equation
\begin{equation}\label{eq:neutral}
    \begin{split}
    F_1(T_F/(T+\Delta))-F_1(-T_F/(T+\Delta))= \\
    = T^2_{F0}/2(T+\Delta)^2 \, ,
    \end{split}
\end{equation}
where $F_1$ is the Fermi integral. For $T \ll T_{F0}$, the susceptibility is inversely proportional to the Fermi temperature, $\chi_c = -1.63 \times 10^{-2}/T_{F0}$, which yields a value of $T_{F0} = 290$~K for $\chi_c(4.2~K) = -5.53 \times 10^{-5}$~emu/g. Taking into account the above relationship between the mean size of crystallites and $T_{F0}$, we obtain $L = 0.91 \times 10^6 \chi_c$~nm = 51~nm in the QTDG sample. Assuming that the proportionality between the crystallite size and susceptibility is also valid at the local level, we can conclude that the NMR spectrum, which reflects the distribution of local susceptibility in the sample, also characterizes the size distribution of crystallites. Hence, we can evaluate the maximum size of crystallites in the sample that should contribute to the left edge of the spectrum in Fig.~\ref{fig:MNR_spectra}. Using the value of susceptibility corresponding to this part of the spectrum, $\chi_c \approx -1.3 \times 10^{-4}$ emu/g, we obtain $L_{max} \approx 120$~nm.

For the independent determination of the characteristic size of crystallites in the QTDG and their size distribution, we investigated the sample surface perpendicular to the texture axis by the method of STM. We used a tunneling microscope with a three-dimensional positioning system for the needle with respect to the sample \cite{troyan12}, which enabled us to obtain images of the surface in different regions of the sample. A typical voltage between the needle and the sample was 10--50~mV, and the tunneling current was about 0.1~nA. After chipping-off the upper part, the sample was mounted, without additional processing, in the tunneling microscope at room temperature on the air, the measurements being made either at room temperature or at 4.2~K. The experiments were carried out as follows. First, we obtained the image of a surface area with a size of 0.1--1 $\mu$m; then, in this area, we chose relatively smooth regions of smaller size for scanning. After detailed investigation of the surface structure of the chosen region, the needle of the microscope was shifted away along the surface to about 0.1--2~mm to study the structure of other regions of the surface.

\begin{figure}[h]
\begin{center}
\includegraphics[width=14cm]{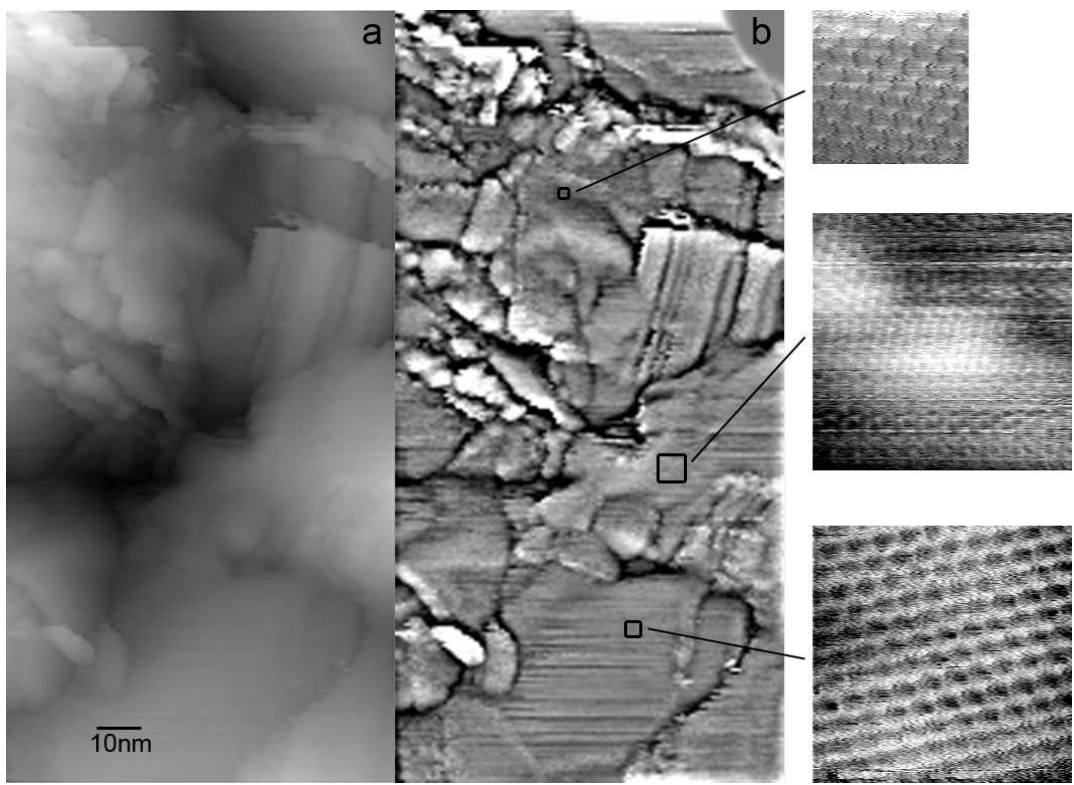}
  \caption{(a) STM image of a 88 $\times$ 176 nm$^2$ surface area of a QTDG sample, $T = 4.2$~K, and (b) the same image after processing by a high-frequency filer. The squares show the regions of individual crystallites whose atomic-resolution images are shown on the right.}
  \label{fig:STM_image}
\end{center}
\end{figure}

\begin{figure}[h]
\begin{center}
\includegraphics[width=13cm]{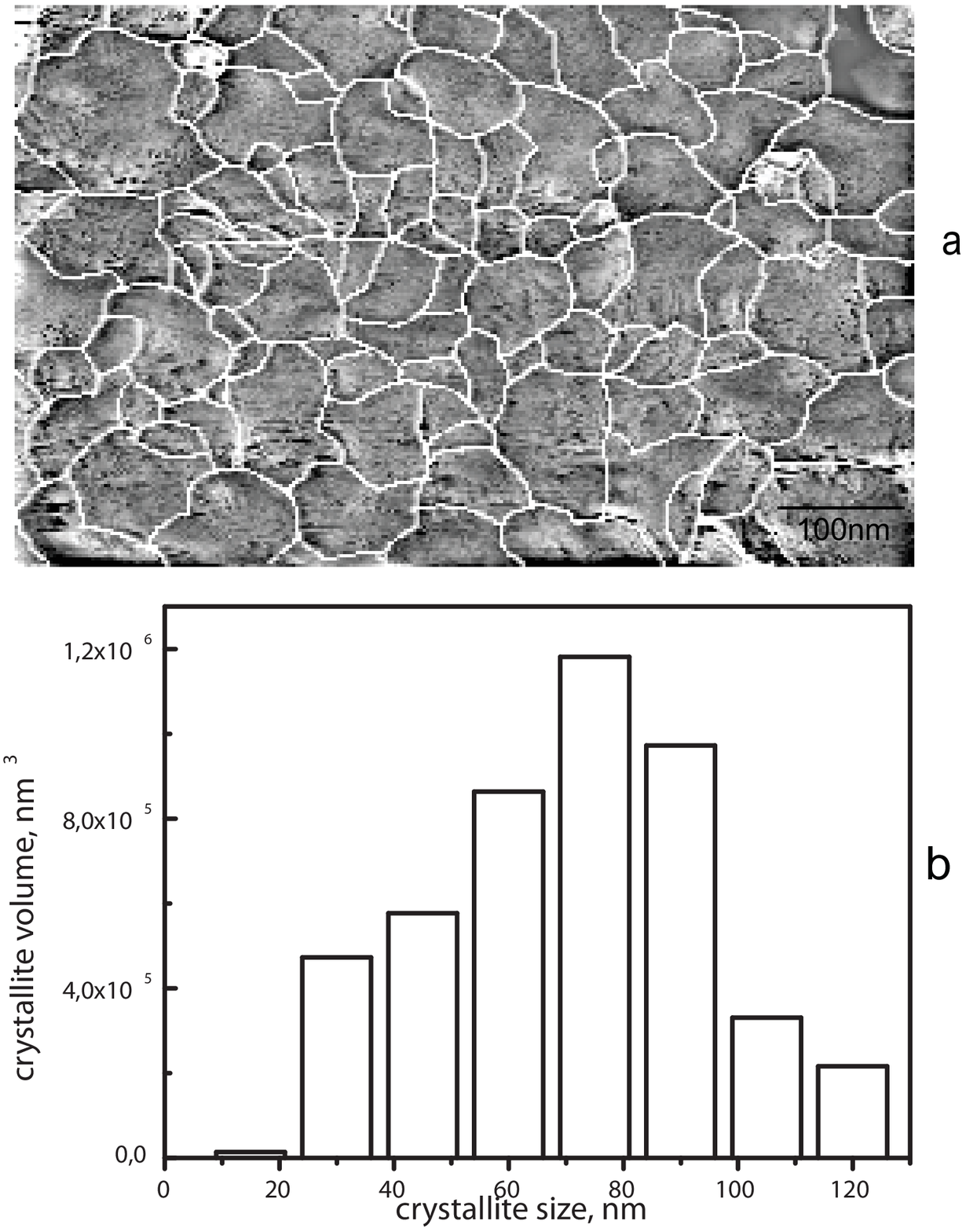}
  \caption{(a) Image of a 711 $\times$ 444 nm$^2$ surface area of a QTDG sample with distinguished boundaries of crystallites after processing by a high-frequency filter, $T = 300$~K, and (b) the size distribution of crystallites in this surface area.}
  \label{fig:size_distr}
\end{center}
\end{figure}

Figure \ref{fig:STM_image}a demonstrates a typical image of a 88 $\times$ 176 nm$^2$ surface area, and Fig.~\ref{fig:STM_image}b shows the result of processing this image by a high-frequency filter. Under this processing, information on the inclination of various regions of the surface is lost; however, the boundaries between these regions become more clearly defined. The images of small surface regions with atomic resolution (see the insets in Fig.~\ref{fig:STM_image}b), which clearly show the honeycomb structure of carbon atoms, made it possible to compare the orientations of the two-dimensional lattice in adjacent regions. Moreover, on the original image, we analyzed height profiles along the line connecting the regions recorded with atomic resolution. It turns out that, as a rule, the regions separated by distinct boundaries have different inclinations and different orientations of carbon planes. This implies that such regions correspond to different crystallites. Using high-frequency filtration, we also processed the images of large surface areas; this allowed us to distinguish a large number of individual crystallites in these areas (Fig.~\ref{fig:size_distr}a) and carry out a statistical analysis of their sizes in the plane of carbon layers. As the crystallite size, we took the geometric mean of the maximum and minimum sizes. Figure \ref{fig:size_distr}b shows the total volume of crystallites (it is the total volume of crystallites of a given size that is proportional to the intensity of the NMR signal at appropriate frequency) as a function of their size, obtained from Fig.~\ref{fig:size_distr}a. In this calculation, we assumed that the thickness of crystallites (the size along the $c$ axis) is independent of their size in the plane of carbon layers and is equal to a mean value of 15~nm evaluated from the width of the (002) X-ray reflex. Otherwise, if, for example, we assume that larger crystallites are thicker, the contribution of these crystallites to the volume will be overstated. According to the figure, the main contribution to the distribution is made by crystallites with a size of about 75~nm, which is noticeably greater than the above value calculated from the magnetic properties. This is likely to be attributed to the fact that the crystallite sizes obtained from STM data are somewhat overstated because the above-described technique for processing STM images does not reveal all the intercrystallite boundaries on the images of large surface areas. Nevertheless, the maximum size of crystallites in the region considered amounts to about 120~nm, which coincides with $L_{max}$ evaluated from the position of the diamagnetic edge of the spectrum. However, we should note that there are crystallites with a size appreciably greater than 120~nm on other images of surface areas.

The value of susceptibility $\chi_c \approx -1.3 \times 10^{-4}$~emu/g at $T = 4.2$~K, which corresponds to the largest crystallites, is the maximum value among known data for non-superconducting systems. Note that, when evaluating this susceptibility, we took into account the demagnetization factor of the sample as a whole, but did not take into account the effect of demagnetization fields of individual crystallites, which may play an appreciable role in such an inhomogeneous system as QTDG (it seems that partly due to this fact the right edge of the spectrum falls into the region of positive shifts). However, this factor will only reduce the local magnetization measured by the NMR method; therefore, the above value of susceptibility for maximum-size crystallites can be considered as a lower bound. In this relation, note that, in \cite{nikolaev09}, the values of susceptibility of $-8 \times 10^{-5}$ emu/g at $T = 4.2$~K for multiwall nanotube carbon columns with the core consisting predominantly of graphite particles with a characteristic size of about 1~$\mu$m were also obtained from the position of the diamagnetic edge of the NMR spectrum. If we take into account that, in \cite{nikolaev09}, demagnetization phenomena were completely neglected, the maximum values of local susceptibility in carbon columns are close to the value observed in our case of QTDG. It seems that such strong diamagnetism is also associated with the absence of correlation between carbon planes in graphite microparticles, which is confirmed by X-ray diffraction data in the paper \cite{bandow96} devoted to the study of various carbon nanostructures.

Thus, the investigation of QTDG by the NMR and STD methods has shown that the observed broad distribution of local magnetization in this system can be explained by the large dispersion of crystallite sizes. The local diamagnetic susceptibility may be more than twice the mean value of $\chi$ over the sample. For the QTDG sample investigated here, it reaches a value of $-1.3 \times 10^{-4}$~emu/g at $T = 4.2$~K, which is greater than the magnetic susceptibility of highly oriented graphite by a factor of four. One may assume that the development of the synthesis methods for QTDG will allow one to obtain a material with much larger crystallites and, hence, with still greater values of diamagnetic susceptibility.

\begin{center}
    \bf{ACKNOWLEDGMENTS}
\end{center}

This work was carried out within the programs "Quantum Mesoscopic and Disordered Structures" and "Fundamentals of the Technology of Nanostructures and Nanomaterials" of the Presidium of the Russian Academy of Sciences.

We are grateful to V.F.~Shamrai and V.P.~Sirotinkin for carrying out X-ray diffraction analysis of the QTDG sample.

\onecolumn

\newpage
\pagestyle{empty} \mbox{} \vspace{1cm}

\newpage
\pagestyle{empty} \mbox{} \vspace{1cm}

%\newpage
\pagestyle{empty} \mbox{} \vspace{1cm}

\end{document}